# Low Energy Neutrino Astronomy in Super-Kamiokande

M. Smy,
*University of California, Irvine, CA 92697, USA*

Super-Kamiokande is sensitive to neutrino interactions between 4 and 100MeV via elastic scattering and inverse beta decay. I will present Super-Kamiokande's ongoing measurements of solar neutrinos and its searches for supernova neutrinos.

## 1. Introduction

In 1987 the star Sanduleak -690 202a (distance: 50kpc) exploded and coincident within 13 sec, 11 neutrino interactions were seen by Kamiokande II [1], 8 neutrino interactions by IMB [1], and 5 neutrino interactions by Baksan. In the same year, Kamiokande observed an excess of events in the solar direction due to solar neutrino-electron elastic scattering [2]. Since then, no extra-terrestrial neutrinos were observed in spite of experimental efforts by many experiments. Super-Kamiokande [3] is a ~50,000 ton water Cherenkov detector optically divided into an inner ~32,000 ton detector viewed by ~11,100 20" photomultiplier tubes and an outer veto region viewed by ~1,800 8" photomultiplier tubes located 2,700m water equivalent underground. Super-Kamiokande has confirmed the solar neutrino-electron elastic scattering measurements of Kamiokande and searched for the diffuse signal due to neutrino interactions from distant supernovae. The dominant supernova neutrino detection mode is the inverse $\beta$ reaction on the free protons in the water molecules.

## 2. Solar Neutrinos

Not only do the solar neutrino flux measurements provide proof that the sun shines via nuclear fusion, they also are evidence for solar neutrino flavor conversion. While this conversion is well described by neutrino oscillation parameters based on KamLAND's measurements of nuclear reactor anti-neutrinos [4], no oscillation signature of the solar neutrino flavor conversion has been observed so far: (i) There should be a characteristic energy dependence of the flavor conversion. Since the higher energy neutrinos (higher energy $^8$B and *hep* neutrinos) undergo complete resonant conversion (due to the MSW [5] effect) in the sun, the flavor changes of the lower energy neutrinos (pp, $^7$Be, pep, the CNO cycle, and lower energy $^8$B) arise from vacuum oscillations averaged out by energy resolution. The resulting energy dependence and the solar neutrino spectrum are shown in Figure 1. The $^8$B flux observed by Super-Kamiokande and SNO and the BOREXINO's $^7$Be flux measurement as well as the pp flux inferred from the charged-current interaction rate on Gallium confirm the oscillation prediction, but no single experiment saw significant energy dependence using the same solar neutrino branch. The transition at around 3 MeV can only be studied with $^8$B neutrinos. (ii) In addition to this spectral signature, $^8$B neutrinos probe matter effects on neutrino oscillations directly by comparing solar neutrinos that pass through the earth at night-time to the day-time solar neutrinos. The analysis of the third phase of Super-Kamiokande solar neutrino data succeeded in reducing systematic uncertainties of the energy measurement of the recoiling electron electrons which will increase its sensitivity to both signatures. It also succeeded in lowering the radioactive backgrounds which limit the energy threshold.

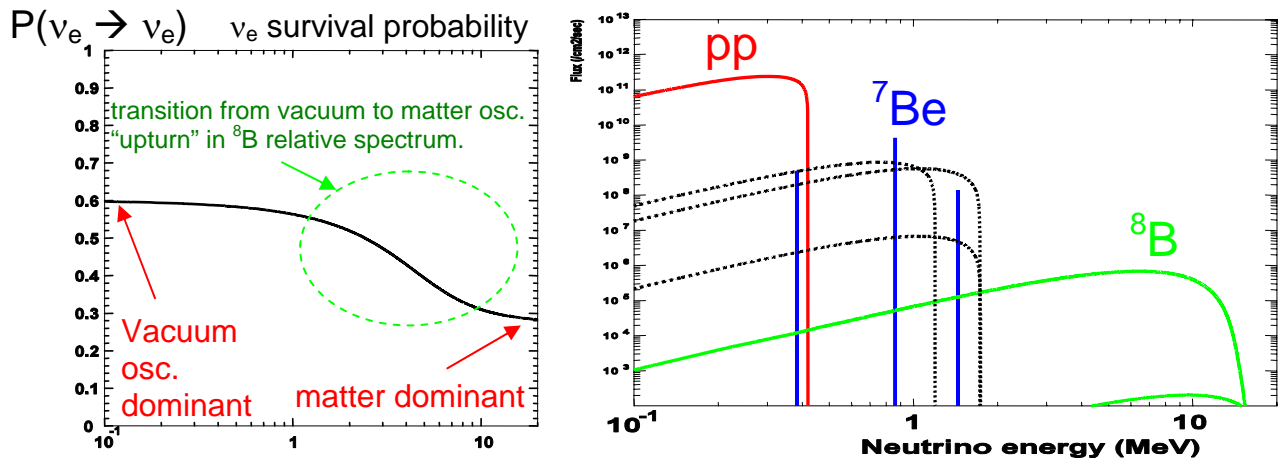

Figure 1: Solar Electron Flavor Neutrino Survival Probability (Left) and Spectrum (Right).

## 2.1. Super-Kamiokande-III Solar Neutrino Results

Above a total recoil electron energy threshold of 6.5 MeV, Super-Kamiokande-III observed solar neutrino – electron elastic scattering during 548 live days. Out of these 548 days, 289 were taken with a lower trigger threshold: The lower energy threshold analysis observed solar neutrino recoil electrons from 4.5 to 6.5 MeV. Figure 2 shows the recoil electron spectrum compared to the expected spectrum (using the 8B spectrum prediction of Winter06 [6]). There is no evidence of spectral distortion. The spectrum and flux is consistent with previous phases of Super-Kamiokande. The flux (above five MeV total recoil electron energy) assuming no oscillation is determined to be $2.32\pm0.04(stat)\pm0.05(syst)\times10^{-6}/cm^2s$; compared to Super-Kamiokande-I's $2.38\pm0.02(stat)\pm0.08(syst)\times10^{-6}/cm^2s$, and Super-Kamiokande-II's $2.41\pm0.05(stat)\pm0.16(syst)\times10^{-6}/cm^2s$. This is the most precise determination of this flux by Super-Kamiokande so far. The day/night ratio $A_{DN}=(day-night)/(½(day+night))$ is measured to be $-0.056\pm0.031(stat)\pm0.013(syst)$. Previously, Super-Kamiokande-I determined this ratio to be $A_{DN}=-0.021\pm0.020(stat)\pm0.013(syst)$, and Super-Kamiokande-II observed a ratio of $A_{DN}=-0.063\pm0.042(stat)\pm0.037(syst)$. The amplitude of the Super-Kamiokande-I day/night variation was fit using the expected shape of the solar zenith angle variation of the data during night-time (shown in Figure 3). The day/night ratio inferred from this fit $-0.018\pm0.016(stat)\pm0.013(syst)$. Applying the same method to all three phases and combining the fits yields $-0.026\pm0.013(stat)$. The systematic uncertainty of this last result is still under study. The fit depends weakly on the solar mass splitting as seen in Figure 3.

The background rate of Super-Kamiokande-III at low energy (<6MeV) is reduced due to a better control of the water flow pattern in the detector. The flow pattern is important because it transports Radon produced by the decay chain of heavy elements in the PMT glass (and the enclosures protecting against PMT implosion chain reactions) from the detector walls to the center. The water flow is controlled by the temperature, flow rate and distribution of the water injected by the purification system during recirculation. While the top inner part of the 22.5 kton fiducial volume becomes very clean, the detector bottom as well as the large radius volume of Super-Kamiokande is contaminated by Radon decays. Removing the bottom and large radii (tight fiducial volume cut), Super-Kamiokande-III measures the solar neutrino rate below 5.5 MeV with a similar precision as Super-Kamiokande-I, although the its livetime is only a quarter of Super-Kamiokande-I's.

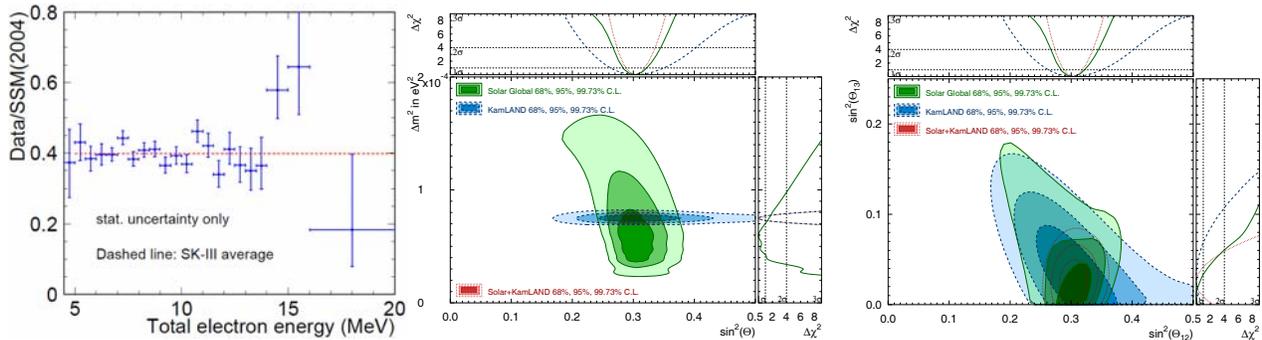

Figure 2: Super-Kamiokande-III Spectral Distortion (Left) and All Solar neutrino/KamLAND Data Three-Flavor Oscillation Analysis (Center and Right).

## 2.2. Oscillation Analysis

A three-flavor neutrino oscillation analysis of the Super-Kamiokande-III data in combination of the data from previous phases, other solar neutrino data [7] (SNO, BOREXINO, Gallex, GNO, SAGE, and the Homestake Chlorine radiochemical experiment), and data from the KamLAND experiment was performed. The analysis determines the solar mass$^2$ splitting (mostly from KamLAND data) and the solar angle $\theta_{12}$ (mostly from SNO and Super-Kamiokande data). There is no indication of a non-zero $\theta_{13}$, a weak bound on this angle is mostly due to solar data (SNO, Super-Kamiokande and BOREXINO) and the different $\theta_{12}/\theta_{13}$ correlation for the solar data (positive correlation) and the KamLAND data (negative correlation). The marginalized allowed areas at 68%, 95% and 99.73% C.L. are shown in Figure 2. Also shown are the marginalized $\Delta\chi^2$ as a function of $\theta_{12}$, $\theta_{13}$, and $\Delta m^2$.

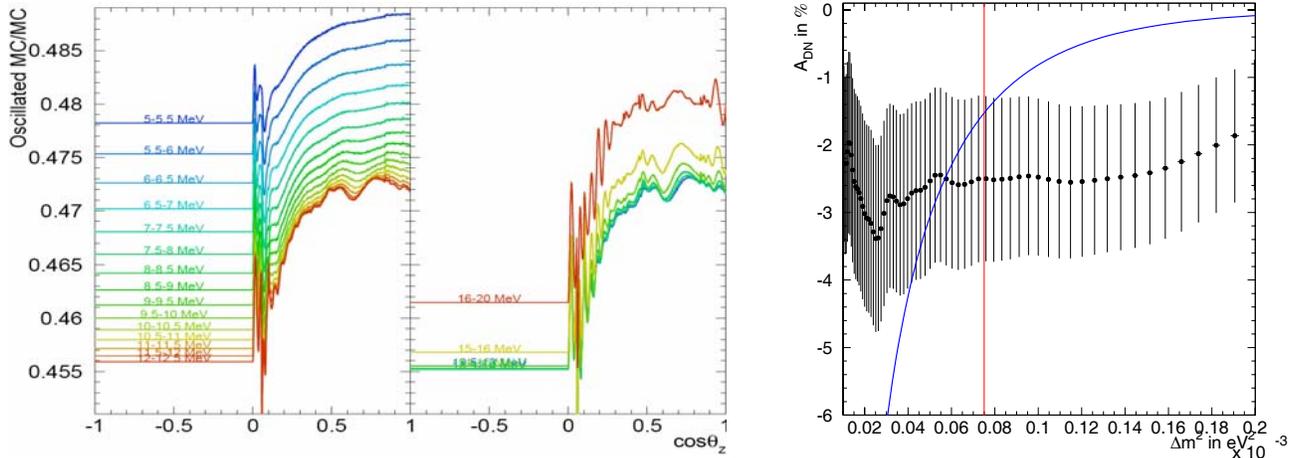

Figure 3: Left Panels: Expected Solar Zenith Angle Variation of the Solar Neutrino Rate in Super-Kamiokande for various Recoil Electron Energy Bins. Right Panel: Combined Super-Kamiokande Data Amplitude Fit to Solar Zenith Angle Variation as a Function of the Solar Mass$^2$ Splitting. The blue function shows the expected amplitude. The red line indicates the preferred value of the solar mass$^2$ splitting.

### 2.3. Super-Kamiokande-IV

Since September 2008, Super-Kamiokande is running with modernized DAQ and electronics. The electronics allows a wider dynamic range in the measured charge and is read out via Ethernet. Because of that, the DAQ is fast enough to record every hit; the resulting data stream is analyzed by an online computer system that finds timing coincidences which are saved as triggers. As a consequence, Super-Kamiokande's energy threshold is now only limited by computing speed and the event reconstruction. The present event reconstruction is able to reconstruct electrons with a total energy of 3 MeV or more. The computing speed limits the energy threshold to ~4.2 MeV which is just a bit below the threshold of Super-Kamiokande-I and III (4.5 MeV). The same water flow control techniques developed during Super-Kamiokande-III result in an obersved solar neutrino elastic scattering peak between 4 to 4.5 MeV total recoil electron energy. To reduce the effective background level, it is useful to compare the hit pattern of background and signal. Since the background is dominated by Radon leading to the β-decay of $^{214}$Bi which has an endpoint of 3.1 MeV, the multiple Coulomb scattering of these electrons is larger than that of solar neutrino recoil electrons which have a harder true spectrum. In order to discriminate, the amount of multiple scattering is estimated by a "directional goodness" which uses a PMT hit-by-PMT hit Hough transformation. If this goodness is close to one, there's little Coulomb scattering, and if it is close to zero, the electron scatters a lot. The solar neutrino candidate sample is then divided into three sub-samples of different amounts of multiple Coulomb scattering. All three sub-samples are fit to a single solar neutrino rate assuming that the signal fraction in each sub-sample is as expected by Monte-Carlo calculations while the background levels of the sub-samples are not constrained. Figure 4 compares the angular distributions for each sub-sample to this fit for the lowest energy bins. It is easy to see the improving signal/background ratio and the "sharpening" of the solar neutrino elastic scattering peak in the lower multiple Coulomb scattering sub-samples. The data is well described by the fit. The statistical uncertainty of this combined fit is up to 10-15% better than a fit to a single sample containing all data in that energy bin. The additional systematic uncertainty of this method is still under study. Of course, the same method can be applied to previous phases of Super-Kamiokande.

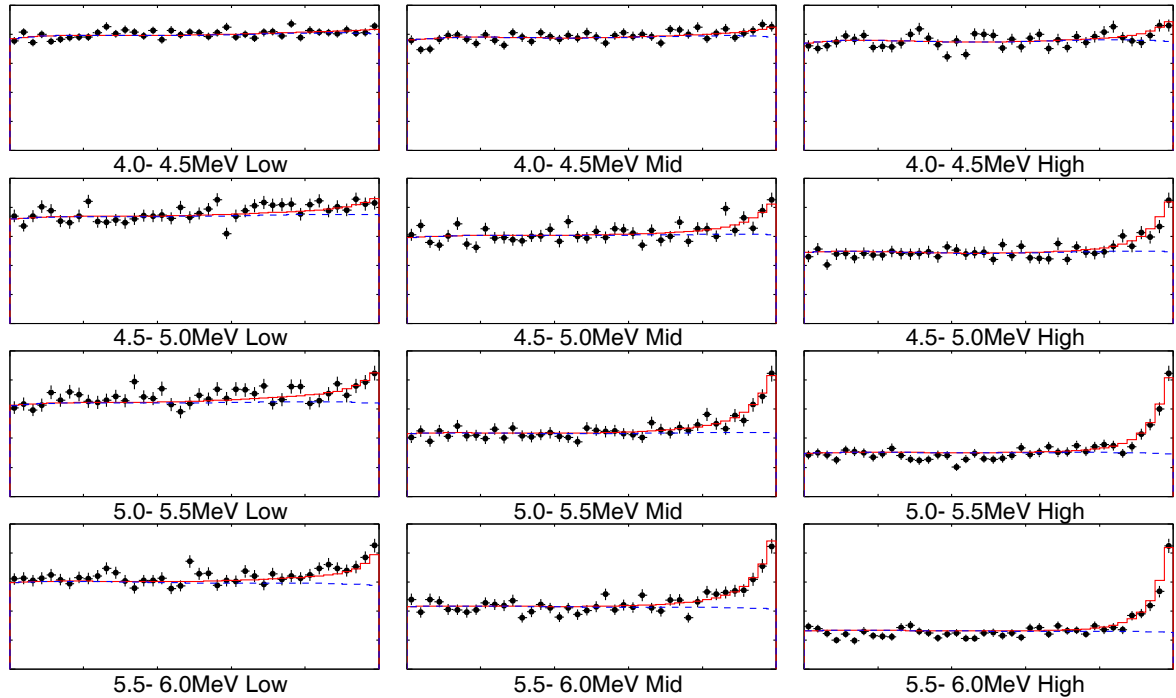

Figure 4: Solar Neutrino Elastic Scattering Peak in Recoil Electron Data in the Large (Left), Medium (Center), and Small (Right) Multiple Coulomb Scattering Data Sub-Sample.

## 3. Search for Diffuse Neutrinos from Distant Supernovae

Massive stars end their life as a supernova, an explosion typically brighter than the host galaxy of the star. Several types of supernovae have been identified; core collapse supernovae happen when the iron core of a sufficiently massive dying star (which cannot produce energy via nuclear fusion) acquires enough mass to collapse gravitationally and force neutronization of protons and electrons into neutrons and neutrinos. If the collapse is halted by the Fermi pressure of the neutrons, a neutron star results, and more than 99% of the binding energy is released in form of neutrinos. The observation of these supernova neutrinos is the unique and only chance to peek into the inside of the dying star. Unfortunately, the small neutrino cross section and present detector masses allow detection of these "neutrino bursts" only if the supernova happens in our own galaxy. An alternative is the continuous measurement of the sum of all core collapse supernovae throughout the universe: a diffuse, isotropic flux of neutrinos. The energy of this signal is somewhat lessened due to red shift. The flux is also quite low because of the large average distance. The signal event rate is dominated by inverse β decay of electron-type anti-neutrinos. Radioactive decays of cosmic muon-induced nuclear spallation products define the lower energy bound of sensitivity: the lifetime of these products can be many seconds, so not all can be tagged by their preceding muons. The upper energy bound is due to atmospheric neutrino background. In particular, charged-current interactions of atmospheric muon neutrinos can produce muons near or below Cherenkov threshold. Those muons remain invisible and their decay electrons are a serious background. Other backgrounds are electronic noise, outside radioactivity (e.g. PMT glass) and solar neutrinos.

The positron spectrum seen by a terrestrial detector depends on the average neutrino emission spectrum, the history of core-collapse supernova, their red-shifts and the inverse β cross section. Even though the supernova neutrino emission spectrum has a rich structure (e.g. due to neutrino – neutrino interactions), the resulting terrestrial positron spectrum is likely to be somewhat insensitive to it since it is an average of many supernovae at different red shifts. Horiuchi, Beacom and Dwek [8] provided a method using an effective two-parameter description of the neutrino emission spectrum which is able to match most calculations. The spectrum is assumed to be Fermi-Dirac and the parameters are the neutrino temperature and neutrino luminosity. Figure 5 shows the resulting positron spectra for an anti-electron neutrino luminosity of $6 \times 10^{52}$ erg.

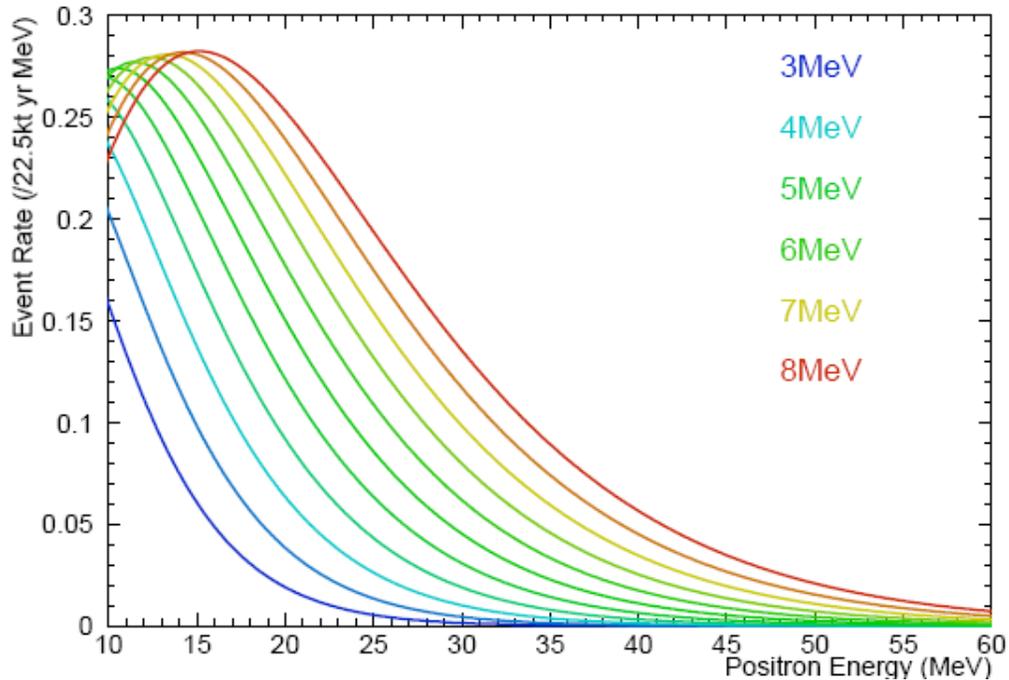

Figure 5. Positron Spectra Resulting from a Fermi-Dirac Effective Eupernova Neutrino Emission Spectrum. The anti-electron neutrino luminosity is $6 \times 10^{52}$ erg.

In 2003, Super-Kamiokande published an upper flux limit of $1.2/cm^2/s$ [9] above 19.3 MeV of neutrino energy (18 MeV total positron energy) based on 1496 days of data collected during the first phase of the experiment (1996-2001). For this search, we greatly increased the efficiency of the analysis, updated the cross section, considered (and removed when possible) additional background sources, improved the statistical analysis and almost doubled the exposure to 2853 days of livetime, 794 days of which (the second phase of the experiment; 2002-2005) were accumulated with reduced inner detector photo-cathode coverage (~5,200 20" photomultiplier tubes). The reduced coverage is due to an accident destroying ~60% of the inner detector photomultiplier tubes. The third phase (2006-2008) of the experiment has the same coverage as the first phase.

In particular, the tagging of spallation products from preceding muons was improved. Previously, a likelihood was formed based on the time difference between muon and candidate, the distance of closest approach, and the "residual charge" (difference between the observed light yield of the muon and the expected light yield of a minimum ionizing track of the observed length). We now reconstruct the dE/dx of these tracks based on the measured times of arrival of Cherenkov photons at the photomultiplier tubes. The peak of the dE/dx profiles of spallation-causing muons correlates with the position of the spallation product. Also, an improved muon fitter categorizes muons into (i) through-going single muons, (ii) stopping muons, (iii) short-track muons ("corner clippers"), and (iv) muon bundles. The likelihood is tuned for each category. As a consequence, the signal efficiency of the spallation cut was improved from 73% to about 90%. The spallation removal efficiency was also improved: after a tighter cut on the likelihood there is no evidence of remaining spallation between 16 MeV and 18 MeV of positron energy. Figure 5 contrasts the correlation of spallation data perpendicular to (top panel) and along the muon track (bottom panel) with the expected random coincidence (a sample formed from subsequent rather than preceding μ's). Table 1 lists the efficiencies.

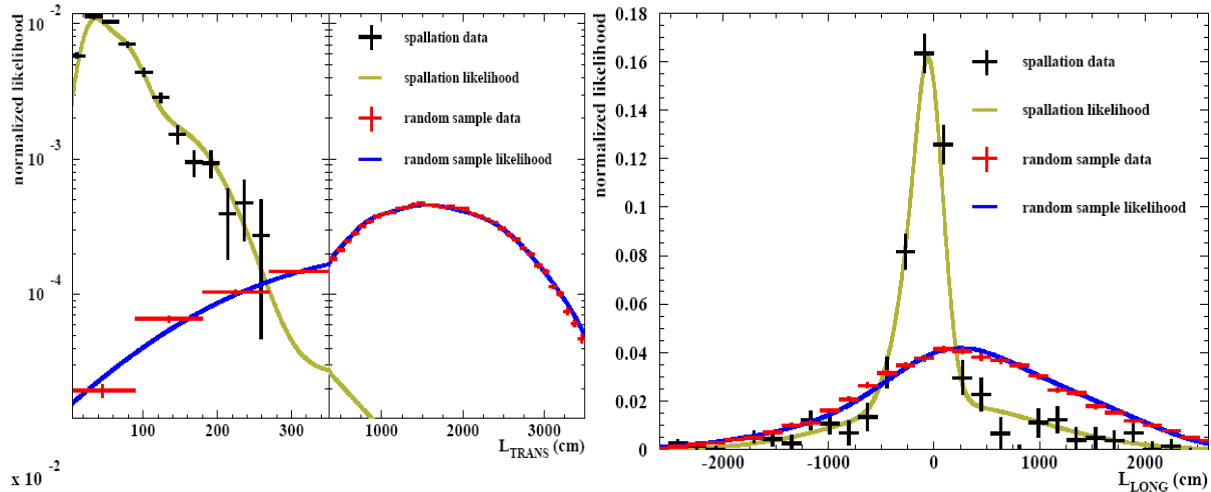

Figure 6. The left (right) panel shows the distance of closest approach $L_{trans}$ (longitudinal distance $L_{long}$) of spallation products to preceding muon tracks in comparison to random coincidence.

Solar neutrinos at 16 MeV form a much larger background than at 18 MeV. Solar neutrino events are due to elastic electron-neutrino scattering, a very forward-peaked process. The removal critically depends on the angular resolution of the recoil electron which is dominated by multiple Coulomb scattering. Reconstructing the amount of multiple Coulomb scattering event by event, the position of a cut placed on the angle between reconstructed recoil electron and solar direction is chosen to minimize the expected signal over the square root of the expected background as a function of energy and the reconstructed amount of multiple Coulomb scattering. The solar cut efficiencies are given in Table 1. Compared to the previously published analysis, several additional cuts remove (atmospheric neutrino) background: candidates with preceding gammas ("pre-activity") or following events ("post-activity") are removed with greater efficiency, charged pions in the sample are identified and rejected by the "sharpness" of their Cherenkov cones, and candidates with correlated outer detector activity are removed (in addition to the normal outer detector veto of events).

| Cut | SK-I/III | | | | SK-II | | | |
|---|---|---|---|---|---|---|---|---|
| Electronic Noise | 99% | | | | 99% | | | |
| Spallation | 81.8%(16-18MeV) | | 90.8%(18-24 MeV) | | 76.2%(17.5-20 MeV) | | 89.2% (20-26 MeV) | |
| Solar ν's | 73.8% (16-17 MeV) | 82.1% (17-18 MeV) | 87.8% (18-19 MeV) | 93% (19-20 MeV) | 74% (17.5-18 MeV) | 82% (18-19 MeV) | 88% (19-20 MeV) | 93% (20-21 MeV) |
| π's (atm. ν's) | 98% | | | | 97% | | | |
| Cher. angle | 95% / 94% | | | | 88% | | | |
| Incoming Event | 98% | | | | 98% | | | |
| Combined Eff. | 78.5%/76.7% | | | | 69.2% | | | |

Table 1. Cut Efficiencies of the analysis.

The published result is based on binned $\chi^2$ test to the energy spectrum of the candidate sample after all cuts were applied. Now we perform an unbinned maximum likelihood fit to three samples: in addition to the signal sample we also fit to a sample which has a large reconstructed Cherenkov angle and to another sample which has a small reconstructed Cherenkov angle. The signal sample contains also decay electrons from atmospheric muon-neutrino charged-current interactions producing invisible muons, atmospheric electron-neutrino charged-current interactions, and atmospheric neutrino neutral-current interactions. To constrain the amount of atmospheric neutral-current background, the large angle sample was added since these interactions often produce several gammas leading to fairly isotropic events. That large angle sample also contains some charged pions and muons from atmospheric neutrino interactions. Many of these however reconstruct at small Cherenkov angles as well, so a fit to all three samples disentangles all three kinds of backgrounds.

Figure 7 shows the data collected by ~11,100 lLeft; phase I and III) and ~5,200 (right) inner detector photomultiplier tubes. Supernova neutrinos would appear in the central panel which shows events with reconstructed Cherenkov cone angles between 38 and 50 degrees. The decay electrons from invisible muons and electrons/positrons from atmospheric $\nu_e$'s will of course also appear in the central panel. Atmospheric neutral-current background mostly reconstruct as isotropic events (right panel; angle above 78 degrees) and below ~20MeV in the central panel. Remaining pions and muons from atmospheric neutrinos show up in the left panel (angle between 20 and 38 degrees) or above ~40MeV in the right panel.

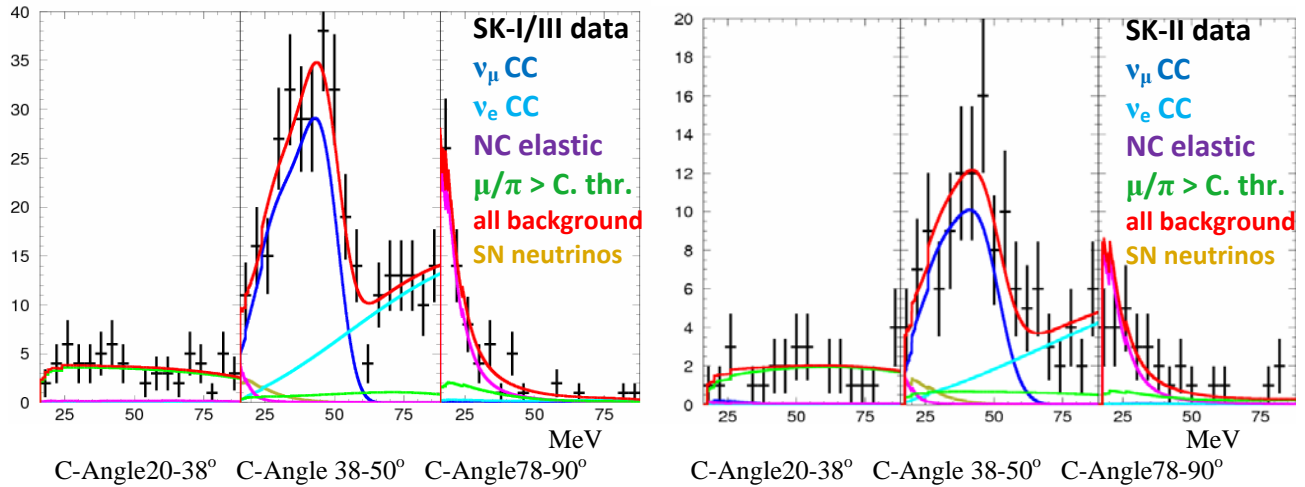

Figure 7. Fit to SK-I/III (left) and SK-II Data (right). The top line shows all backgrounds combined. In the low angle region, all background is due to charged pion and muon production from atmospheric neutrinos. The high angle region is dominated by atmospheric neutrino neutral current interactions with a small pion/muon component. The central region is dominated by invisible muon production from atmospheric neutrinos. Above ~60 MeV, atmospheric electron-neutrinos dominate. Below 18MeV, neutral current interactions start to become important. The central plot also shows the best fit relic contribution.

The data in Figure 7 are well described by just atmospheric neutrino background, so no significant excess due to supernova neutrinos was found. The likelihood of the fit is used to place a 90% upper limit on the supernova neutrino flux. To obtain this limit, the likelihood was marginalized over all background rate parameters to obtain the likelihood as a function of only the supernova event rate. The 90% limit is defined as the rate below which 90% of the area underneath that likelihood is contained. Typical likelihoods of the first, second, and third phase of the experiment as well as the combined likelihood are shown in Figure 4. In this case, the 90% C.L. limit is about <5events/year for positron energies above 16 MeV which corresponds to a flux of about 2.7/cm$^2$/s above 17.3 MeV. Figure 8 shows the 90% C.L. event rate limit as a function of temperature (dashed line) and the 90% C.L. excluded area of event rate versus neutrino temperature. Table 2 lists the typical supernova anti-electron neutrino luminosity limit for each temperature assuming a Fermi-Dirac emission spectrum.

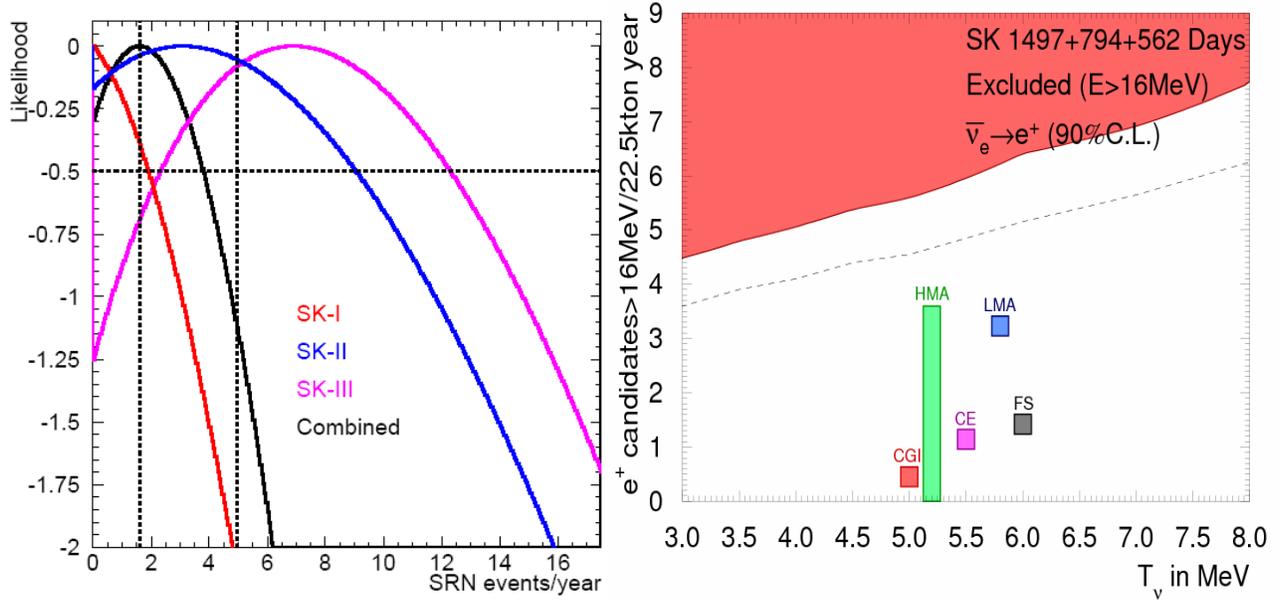

Figure 8. Left Panel: Example of Likelihood (T=6MeV) as a Function of Supernova ν Event Rate. The intersection with the dotted horizontal line illustrates the 1σ range of the fit. The right (left) dotted vertical line shows the 90% C.L. upper limit (best fit) of the supernova event rate. Right Panel: Excluded Area of Supernova Event rate Versus Neutrino Temperature. The dashed line gives the event rate limit given a particular neutrino temperature. The dashed line shows the 90 % C.L. limit obtained for a fixed neutrino temperature. Overlaid are various theoretical models [10].

| T [MeV]        | 3   | 3.5 | 4   | 4.5 | 5.0 | 5.5 | 6.0 | 6.5  | 7.0  | 7.5  | 8.0  |
|----------------|-----|-----|-----|-----|-----|-----|-----|------|------|------|------|
| L [$10^{53}$ erg] | 9.0 | 4.8 | 2.9 | 2.1 | 1.5 | 1.3 | 1.1 | 0.92 | 0.82 | 0.75 | 0.70 |

Table 2. 90% C.L. Upper Luminosity Limit for a Given Anti-electron Neutrino Temperature T.